\documentclass[12pt]{article}
\usepackage{amssymb}
\begin{document}



\def\a{\alpha}
\def\b{\beta}
\def\d{\delta}
\def\e{\epsilon}
\def\g{\gamma}
\def\h{\mathfrak{h}}
\def\k{\kappa}
\def\l{\lambda}
\def\o{\omega}
\def\p{\wp}
\def\r{\rho}
\def\t{\tau}
\def\s{\sigma}
\def\z{\zeta}
\def\x{\xi}
 \def\A{{\cal{A}}}
 \def\B{{\cal{B}}}
 \def\C{{\cal{C}}}
 \def\D{{\cal{D}}}
\def\G{\Gamma}
\def\K{{\cal{K}}}
\def\O{\Omega}
\def\R{\bar{R}}
\def\T{{\cal{T}}}
\def\L{\Lambda}
\def\f{E_{\tau,\eta}(sl_2)}
\def\E{E_{\tau,\eta}(sl_n)}
\def\Zb{\mathbb{Z}}
\def\Cb{\mathbb{C}}

\def\R{\overline{R}}

\def\beq{\begin{equation}}
\def\eeq{\end{equation}}
\def\bea{\begin{eqnarray}}
\def\eea{\end{eqnarray}}
\def\ba{\begin{array}}
\def\ea{\end{array}}
\def\no{\nonumber}
\def\le{\langle}
\def\re{\rangle}
\def\lt{\left}
\def\rt{\right}

\newtheorem{Theorem}{Theorem}
\newtheorem{Definition}{Definition}
\newtheorem{Proposition}{Proposition}
\newtheorem{Lemma}{Lemma}
\newtheorem{Corollary}{Corollary}
\newcommand{\proof}[1]{{\bf Proof. }
        #1\begin{flushright}$\Box$\end{flushright}}

\baselineskip=20pt

\newfont{\elevenmib}{cmmib10 scaled\magstep1}
\newcommand{\preprint}{
   \begin{flushleft}
   \end{flushleft}\vspace{-1.3cm}
   \begin{flushright}\normalsize
     {\tt hep-th/0703222} \\ March 2007
   \end{flushright}}
\newcommand{\Title}[1]{{\baselineskip=26pt
   \begin{center} \Large \bf #1 \\ \ \\ \end{center}}}
\newcommand{\Author}{\begin{center}
   \large \bf
Wen-Li Yang${}^{a,b}$
 ~and~ Yao-Zhong Zhang ${}^b$
 \end{center}}
\newcommand{\Address}{\begin{center}

     ${}^a$ Institute of Modern Physics, Northwest University,
     Xian 710069, P.R. China\\
     ${}^b$ Department of Mathematics, University of Queensland, Brisbane, QLD 4072,
     Australia\\
     E-mail: wenli@maths.uq.edu.au, yzz@maths.uq.edu.au
   \end{center}}
\newcommand{\Accepted}[1]{\begin{center}
   {\large \sf #1}\\ \vspace{1mm}{\small \sf Accepted for Publication}
   \end{center}}

\preprint
\thispagestyle{empty}
\bigskip\bigskip\bigskip

\Title{On the second reference state and complete eigenstates of
the open XXZ  chain} \Author

\Address
\vspace{1cm}

\begin{abstract}
The second reference state of the open XXZ spin chain with
non-diagonal boundary terms is studied. The associated Bethe
states exactly yield the second set of eigenvalues proposed
recently by functional Bethe Ansatz. In the quasi-classical limit,
two sets of Bethe states give the complete eigenstates of the
associated Gaudin model.

\vspace{1truecm} \noindent {\it PACS:} 03.65.Fd; 04.20.Jb;
05.30.-d; 75.10.Jm

\noindent {\it Keywords}: Algebraic Bethe ansatz; Gaudin model.
\end{abstract}
\newpage
\section{Introduction}
\label{intro} \setcounter{equation}{0}

The open XXZ quantum spin chain has played a fundamental role in
the study of quantum integrable systems with various boundary
interactions, which appeared in statistical mechanics, condensed
matter and quantum field theory. Although the special case of
diagonal boundary terms was solved long ago
\cite{Gau71,Alc87,Skl88}, Bethe Ansatz solutions for non-diagonal
boundary terms where the boundary parameters obey some constraints
have been proposed only recently by various approaches
\cite{Nep04,Cao03,Yan04-1,Gie05,Doi06,Yan06}. It is found that in
order to obtain the complete spectrum of the model two sets of
Bethe Ansatz equations and consequently two sets of eigenvalues
are {\it needed} \cite{Nep03,Yan06}, in contrast with the diagonal
boundary case \cite{Skl88}. This suggests that in the framework of
algebraic Bethe Ansatz there should exist two reference states (or
pseudo-vacuum states) corresponding to the two sets of Bethe
Ansatz equations and eigenvalues. However, to our knowledge only
one reference has been constructed so far \cite{Cao03,Yan04-1}.
Moreover, the explicit expressions of the complete reference
states and associated Bethe states are of great {\it importance}
for investigating correlation functions of the model \cite{Kor93}.

In this letter we  study the second reference state of the open
XXZ spin chain with non-diagonal boundary terms and construct the
{\it complete} eigenstates of the model in the framework of
algebraic Bethe Ansatz. In the quasi-classical limit, they give
the complete eigenstates of the associated Gaudin model.


\section{ The inhomogeneous spin-$\frac{1}{2}$ XXZ open chain}
\label{XXZ} \setcounter{equation}{0}

Throughout, $V$ denotes a two-dimensional linear space and
$\s^{\pm},\,\s^z$ are the usual Pauli matrices which realize the
spin-$\frac{1}{2}$ representation of the Lie algebra $sl(2)$ on
$V$. The spin-$\frac{1}{2}$ XXZ chain can be constructed from the
well-known six-vertex model R-matrix $\R(u)\in {\rm End}(V\otimes
V)$ \cite{Kor93} given by \bea
\R(u)=\lt(\begin{array}{llll}1&&&\\&b(u)&c(u)&\\
&c(u)&b(u)&\\&&&1\end{array}\rt).\label{r-matrix}\eea The
coefficient functions read: $b(u)=\frac{\sin u}{\sin(u+\eta)}$,
$c(u)=\frac{\sin\eta}{\sin(u+\eta)}$. Here we assume  $\eta$ being
a generic complex number. The R-matrix satisfies the quantum
Yang-Baxter equation (QYBE), \bea
R_{12}(u_1-u_2)R_{13}(u_1-u_3)R_{23}(u_2-u_3)
=R_{23}(u_2-u_3)R_{13}(u_1-u_3)R_{12}(u_1-u_2),\label{QYB}\eea and
the unitarity, crossing-unitarity and quasi-classical properties
\cite{Yan04-1}. We adopt the standard notations: for any matrix
$A\in {\rm End}(V)$ , $A_j$ is an embedding operator in the tensor
space $V\otimes V\otimes\cdots$, which acts as $A$ on the $j$-th
space and as identity on the other factor spaces; $R_{ij}(u)$ is
an embedding operator of R-matrix in the tensor space, which acts
as identity on the factor spaces except for the $i$-th and $j$-th
ones.

One introduces the ``row-to-row" monodromy matrix $T(u)$, which is
a $2\times 2$ matrix with elements being operators acting on
$V^{\otimes N}$, where $N=2M$ ($M$ being a positive integer),\bea
T_0(u)=\R_{01}(u+z_1)\R_{02}(u+z_2)\cdots
\R_{0N}(u+z_N).\label{Mon-V}\eea Here $\{z_j|j=1,\cdots,N\}$ are
arbitrary free complex parameters which are usually called
inhomogeneous parameters.

Integrable open chain can be constructed as follows \cite{Skl88}.
Let us introduce a pair of K-matrices $K^-(u)$ and $K^+(u)$. The
former satisfies the reflection equation (RE)
 \bea &&\R_{12}(u_1-u_2)K^-_1(u_1)\R_{21}(u_1+u_2)K^-_2(u_2)\no\\
 &&~~~~~~=
K^-_2(u_2)\R_{12}(u_1+u_2)K^-_1(u_1)\R_{21}(u_1-u_2),\label{RE-V}\eea
and the latter  satisfies the dual RE \bea
&&\R_{12}(u_2-u_1)K^+_1(u_1)\R_{21}(-u_1-u_2-2\eta)K^+_2(u_2)\no\\
&&~~~~~~=
K^+_2(u_2)\R_{12}(-u_1-u_2-2\eta)K^+_1(u_1)\R_{21}(u_2-u_1).
\label{DRE-V}\eea For open spin-chains, instead of the standard
``row-to-row" monodromy matrix $T(u)$ (\ref{Mon-V}), one needs to
introduce the
 ``double-row" monodromy matrix $\mathbb{T}(u)=T(u)K^-(u)T^{-1}(-u)$.
Then the {\it double-row transfer matrix\/}  is given by \bea
\t(u)=tr(K^+(u)\mathbb{T}(u)).\label{trans}\eea The QYBE and
(dual) REs lead to that the transfer matrices with different
spectral parameters commute with each other \cite{Skl88}:
$[\t(u),\t(v)]=0$. This ensures the integrability of the
inhomogeneous spin-$\frac{1}{2}$ XXZ chain with open boundary.

In this paper, we will consider a {\it generic\/} K-matrix
$K^{-}(u)$ which is a generic solution to the RE (\ref{RE-V})
associated the six-vertex model R-matrix  \cite{Veg93,Gho94}
\bea K^-(u)=\lt(\begin{array}{ll}k_1^1(u)&k^1_2(u)\\
k^2_1(u)&k^2_2(u)\end{array}\rt)\equiv K(u).\label{K-matrix}\eea
The coefficient functions are \bea && k^1_1(u)=
\frac{2\cos(\l_1-\l_2) -\cos(\l_1+\l_2+2\xi)e^{-2iu}}
{4\sin(\l_1+\xi+u)
\sin(\l_2+\xi+u)},\no\\
&&k^1_2(u)=\frac{-i\sin(2u)e^{-i(\l_1+\l_2)} e^{-iu}}
{2\sin(\l_1+\xi+u) \sin(\l_2+\xi+u)},\no\\
&&k^2_1(u)=\frac{i\sin(2u)e^{i(\l_1+\l_2)} e^{-iu}}
{2\sin(\l_1+\xi+u) \sin(\l_2+\xi+u)}, \no\\
&& k^2_2(u)=\frac{2\cos(\l_1-\l_2)e^{-2iu}- \cos(\l_1+\l_2+2\xi)}
{4\sin(\l_1+\xi+u)\sin(\l_2+\xi+u)}.\label{K-matrix-2-1} \eea At
the same time, we introduce  the corresponding {\it dual\/}
K-matrix $K^+(u)$ which is a generic solution to the dual
reflection equation (\ref{DRE-V}) with a particular choice of the
free boundary parameters with respect to $K^-(u)$:
\bea K^+(u)=\lt(\begin{array}{ll}{k^+}_1^1(u)&{k^+}^1_2(u)\\
{k^+}^2_1(u)&{k^+}^2_2(u)\end{array}\rt).\label{DK-matrix}\eea The
matrix elements are \bea && {k^+}^1_1(u)=
\frac{2\cos(\l_1-\l_2)e^{-i\eta}
-\cos(\l_1+\l_2+2\bar{\xi})e^{2iu+i\eta}}
{4\sin(\l_1+\bar{\xi}-u-\eta)
\sin(\l_2+\bar{\xi}-u-\eta)},\no\\
&&{k^+}^1_2(u)=\frac{i\sin(2u+2\eta)e^{-i(\l_1+\l_2)}
e^{iu-i\eta}} {2\sin(\l_1+\bar{\xi}-u-\eta)
\sin(\l_2+\bar{\xi}-u-\eta)},
\no\\
&&{k^+}^2_1(u)=\frac{-i\sin(2u+2\eta)e^{i(\l_1+\l_2)}
e^{iu+i\eta}}
{2\sin(\l_1+\bar{\xi}-u-\eta) \sin(\l_2+\bar{\xi}-u-\eta)}, \no\\
&& {k^+}^2_2(u)=\frac{2\cos(\l_1-\l_2)e^{2iu+i\eta}-
\cos(\l_1+\l_2+2\bar{\xi})e^{-i\eta}}
{4\sin(\l_1+\bar{\xi}-u-\eta)\sin(\l_2+\bar{\xi}-u-\eta)}.\label{K-matrix-6}
\eea The K-matrices depend on four free boundary parameters
$\{\l_1,\,\l_2,\,\xi,\,\bar{\xi}\}$ which obey the constrain
conditions in \cite{Nep04,Cao03,Yan06}. It is very convenient to
introduce a vector $\l=\sum_{k=1}^2\l_k\e_k$ associated with the
boundary parameters $\{\l_i\}$, where $\{\e_i,\,i=1,2\}$ form the
orthonormal basis of $V$ such that $\langle
\e_i,\e_j\rangle=\d_{ij}$.


\section{Vertex-face correspondence}
\label{BAE} \setcounter{equation}{0}

Let us briefly review the face-type R-matrix associated with the
six-vertex model. Set $\hat{i}=\e_i-\overline{\e},~~\overline{\e}=
\frac{1}{2}\sum_{k=1}^{2}\e_k$, for $i=1,2$. For a generic $m\in
V$, define \bea m_i=\langle m,\e_i\rangle,
~~m_{ij}=m_i-m_j=\langle m,\e_i-\e_j\rangle,~~i,j=1,2.
\label{Def1}\eea Let $R(u,m)\in {\rm End}(V\otimes V)$ be the
R-matrix of the six-vertex SOS model, which is trigonometric limit
of the eight-vertex SOS model \cite{Bax82} given by \bea
&&R(u,m)\hspace{-0.1cm}=\hspace{-0.1cm}
\sum_{i=1}^{2}R^{ii}_{ii}(u,m)E_{ii}\hspace{-0.1cm}\otimes\hspace{-0.1cm}
E_{ii}\hspace{-0.1cm}+\hspace{-0.1cm}\sum_{i\ne
j}^2\lt\{R^{ij}_{ij}(u,m)E_{ii}\hspace{-0.1cm}\otimes\hspace{-0.1cm}
E_{jj}\hspace{-0.1cm}+\hspace{-0.1cm}
R^{ji}_{ij}(u,m)E_{ji}\hspace{-0.1cm}\otimes\hspace{-0.1cm}
E_{ij}\rt\}, \label{R-matrix} \eea where $E_{ij}$ is the matrix
with elements $(E_{ij})^l_k=\d_{jk}\d_{il}$. The coefficient
functions are \bea &&R^{ii}_{ii}(u,\l)=1,~~
R^{ij}_{ij}(u,\l)=\frac{\sin u\sin(m_{ij}-\eta)}
{\sin(u+\eta)\sin(m_{ij})},~~i\neq j,\label{Elements1}\\
&& R^{ji}_{ij}(u,m)=\frac{\sin\eta\sin(u+m_{ij})}
{\sin(u+\eta)\sin(m_{ij})},~~i\neq j,\label{Elements2}\eea  and
$m_{ij}$ is defined in (\ref{Def1}). The R-matrix satisfies the
dynamical (modified) quantum Yang-Baxter equation (or
star-triangle equation) \cite{Bax82}.

Define the following functions: $\theta^{(1)}(u)=e^{-iu}$, $
\theta^{(2)}(u)=1$. Let us introduce two intertwiners which are
$2$-component  column vectors $\phi_{m,m-\eta\hat{j}}(u)$ labelled
by $\hat{1},\,\hat{2}$. The $k$-th element of
$\phi_{m,m-\eta\hat{j}}(u)$ is given by \bea
\phi^{(k)}_{m,m-\eta\hat{j}}(u)=\theta^{(k)}(u+2m_j).\label{Intvect}\eea
Explicitly, \bea \phi_{m,m-\eta\hat{1}}(u)=
\lt(\begin{array}{c}e^{-i(u+2m_1)}\\1\end{array}\rt),\qquad
\phi_{m,m-\eta\hat{2}}(u)=
\lt(\begin{array}{c}e^{-i(u+2m_2)}\\1\end{array}\rt).\eea
Obviously, the two intertwiner vectors $\phi_{m,m-\eta\hat{i}}(u)$
are linearly {\it independent} for a generic $m\in V$.

 Using the
intertwiner vectors, one can derive the following face-vertex
correspondence relation \cite{Cao03}\bea &&\R_{12}(u_1-u_2)
\phi_{m,m-\eta\hat{\imath}}(u_1)\otimes
\phi_{m-\eta\hat{\imath},m-\eta(\hat{\imath}+\hat{\jmath})}(u_2)
\no\\&&~~~~~~= \sum_{k,l}R(u_1-u_2,m)^{kl}_{ij}
\phi_{m-\eta\hat{l},m-\eta(\hat{l}+\hat{k})}(u_1)\otimes
\phi_{m,m-\eta\hat{l}}(u_2). \label{Face-vertex}\eea Then the QYBE
(\ref{QYB}) of the vertex-type R-matrix $\R(u)$ is equivalent to
the dynamical Yang-Baxter equation of the SOS R-matrix $R(u,m)$.
For a generic $m$, we can introduce other types of intertwiners
$\bar{\phi},~\tilde{\phi}$ satisfying the conditions, \bea
&&\sum_{k=1}^2\bar{\phi}^{(k)}_{m,m-\eta\hat{\mu}}(u)
~\phi^{(k)}_{m,m-\eta\hat{\nu}}(u)=\d_{\mu\nu},\quad
\sum_{k=1}^2\tilde{\phi}^{(k)}_{m+\eta\hat{\mu},m}(u)
~\phi^{(k)}_{m+\eta\hat{\nu},m}(u)=\d_{\mu\nu}.\label{Int2}\eea
One may verify that the K-matrices $K^{\pm}(u)$ given by
(\ref{K-matrix}) and (\ref{DK-matrix}) can be expressed in terms
of the intertwiners and {\it diagonal\/} matrices $\K(\l|u)$ and
$\tilde{\K}(\l|u)$ as follows \bea &&K^-(u)^s_t=
\sum_{i,j}\phi^{(s)}_{\l-\eta(\hat{\imath}-\hat{\jmath}),
~\l-\eta\hat{\imath}}(u)
\K(\l|u)^j_i\bar{\phi}^{(t)}_{\l,~\l-\eta\hat{\imath}}(-u),\label{K-F-1}\\
&&K^+(u)^s_t= \sum_{i,j}
\phi^{(s)}_{\l,~\l-\eta\hat{\jmath}}(-u)\tilde{\K}(\l|u)^j_i
\tilde{\phi}^{(t)}_{\l-\eta(\hat{\jmath}-\hat{\imath}),
~\l-\eta\hat{\jmath}}(u).\label{K-F-2}\eea Here the two {\it
diagonal\/} matrices $\K(\l|u)$ and $\tilde{\K}(\l|u)$ are given
by \bea &&\K(\l|u)\equiv{\rm Diag}(k(\l|u)_1,\,k(\l|u)_2)={\rm
Diag}(\frac{\sin(\l_1+\xi-u)}{\sin(\l_1+\xi+u)},\,
\frac{\sin(\l_2+\xi-u)}{\sin(\l_2+\xi+u)}),\label{K-F-3}\\
&&\tilde{\K}(\l|u)\equiv{\rm
Diag}(\tilde{k}(\l|u)_1,\,\tilde{k}(\l|u)_2)\no\\
&&~~~~~~~~~={\rm
Diag}(\frac{\sin(\l_{12}\hspace{-0.1cm}-\hspace{-0.1cm}
\eta)\sin(\l_1\hspace{-0.1cm}+\hspace{-0.1cm}\xi+\hspace{-0.1cm}u
\hspace{-0.1cm}+\hspace{-0.1cm}\eta)}
{\sin\l_{12}\sin(\l_1+\xi-u-\eta)},\,
\frac{\sin(\l_{12}\hspace{-0.1cm}+\hspace{-0.1cm}
\eta)\sin(\l_2\hspace{-0.1cm}+\hspace{-0.1cm}\xi\hspace{-0.1cm}
+\hspace{-0.1cm}u\hspace{-0.1cm}+\hspace{-0.1cm}\eta)}
{\sin\l_{12}\sin(\l_2+\xi-u-\eta)}).\label{K-F-4} \eea Although
the K-matrices $K^{\pm}(u)$ given by (\ref{K-matrix}) and
(\ref{DK-matrix}) are generally non-diagonal (in the vertex form),
after the face-vertex transformations (\ref{K-F-1}) and
(\ref{K-F-2}), the face type counterparts $\K(\l|u)$ and
$\tilde{\K}(\l|u)$ {\it simultaneously\/} become diagonal. This
fact enables us to apply the generalized algebraic Bethe ansatz
method developed in \cite{Yan04} for SOS type integrable models to
diagonalize the transfer matrices $\t(u)$ (\ref{trans}).

The decomposition of $K^+(u)$ (\ref{K-F-2}) and the diagonal
property (\ref{K-F-4}) lead to the recasting of the transfer
matrix $\t(u)$ (\ref{trans}) in the following face type form \bea
&&\t(u)=tr(K^+(u)\mathbb{T}(u))
=\sum_{\mu,\nu}\tilde{\K}(\l|u)_{\nu}^{\mu}\T(\l|u)^{\nu}_{\mu}=
\sum_{\mu}\tilde{k}(\l|u)_{\mu}\T(\l|u)^{\mu}_{\mu}.
\label{De1}\eea Here we have introduced the face-type double-row
monodromy matrix $\T(m|u)$ \bea \T(m|u)^{\nu}_{\mu}
=\sum_{i,j}\tilde{\phi}^{(j)}_{m-\eta(\hat{\mu}-\hat{\nu}),
m-\eta\hat{\mu}}(u)~\mathbb{T}(u)^j_i\phi^{(i)}_{m,
m-\eta\hat{\mu}}(-u).\label{Mon-F} \eea This face-type double-row
monodromy matrix can  be expressed in terms of the face type
R-matrix $R(m|u)$ (\ref{R-matrix}) and K-matrix $\K(\l|u)$
(\ref{K-F-3}) (for the details, see equation (4.19) of
\cite{Yan04}).

As in \cite{Yan04-1}, let us introduce operators: \bea
&&\A(m|u)\equiv \hspace{-0.1truecm}\A^{(1)}(m|u)=
\hspace{-0.1truecm}\T(m|u)^1_1,~~\B(m|u)=
\hspace{-0.1truecm}
\frac{\T(m|u)^1_2}{\sin(m_{12})},
~~\C(m|u)=
\hspace{-0.1truecm}
\frac{\T(m|u)^2_1}{\sin(m_{21})},\label{Def-AB}\\
&&\D(m|u)\equiv \hspace{-0.1truecm}\D^{(1)}(m|u)=
\hspace{-0.1truecm}\frac{\sin(m_{12}+\eta)}{\sin(m_{12})}
\{\T(m|u)^2_2\hspace{-0.1truecm}-\hspace{-0.1truecm}
R(2u,m+\eta\hat{1})^{2\,1}_{1\,2}\A^{(1)}(m|u)\}.
\label{Def-D}\eea We remark that the transfer matrix $\t(u)$
(\ref{trans}) can be expressed in terms of the operators
$\A^{(1)}$ and $\D^{(1)}$. It was found in \cite{Yan04-1} that one
can construct a reference state, denoted by
$|\O^{(1)}(\l)\rangle$, \bea
\hspace{-1.2truecm}|\O^{(1)}(\l)\rangle
&=&\phi_{\l-(N-1)\eta\hat{1},\l-N\eta\hat{1}}(-z_1)\otimes
\phi_{\l-(N-2)\eta\hat{1},\l-(N-1)\eta\hat{1}}(-z_{2})\cdots\otimes
\phi_{\l,\l-\eta\hat{1}}(-z_N),\label{Vac} \eea in the sense that
the state is common eigenstate of the operators $\A^{(1)}$ and
$\D^{(1)}$ and is annihilated by $\C$ (c.f.(\ref{Ann-1})). The
associated Bethe states can be constructed by applying the
``creation operator" $\B$ on the corresponding reference state
\bea |v_1,\cdots,v_M\rangle^{(1)}=\B(\l-2\eta\hat{1}|v_1)
\B(\l-4\eta\hat{1}|v_2)\cdots\B(\l-2M\eta\hat{1}|v_M)|\O^{(1)}(\l)\rangle.
\label{Bethe-state}\eea If the parameters $\{v_k\}$ satisfy the
following Bethe Ansatz equations, \bea &&\hspace{-0.1cm}\frac
{\sin(\l_2+\xi+v_{\a})\sin(\l_2+\bar\xi-v_{\a})
\sin(\l_1+\bar\xi+v_{\a})\sin(\l_1+\xi-v_{\a})}
{\sin(\l_2\hspace{-0.1cm}+\hspace{-0.1cm}
\bar\xi\hspace{-0.1cm}+\hspace{-0.1cm}v_{\a}
\hspace{-0.1cm}+\hspace{-0.1cm}\eta)
\sin(\l_2\hspace{-0.1cm}+\hspace{-0.1cm}\xi\hspace{-0.1cm}-\hspace{-0.1cm}v_{\a}
\hspace{-0.1cm}-\hspace{-0.1cm}\eta)
\sin(\l_1\hspace{-0.1cm}+\hspace{-0.1cm}\xi\hspace{-0.1cm}+\hspace{-0.1cm}
v_{\a}\hspace{-0.1cm}+\hspace{-0.1cm}\eta)
\sin(\l_1\hspace{-0.1cm}+\hspace{-0.1cm}\bar\xi\hspace{-0.1cm}-\hspace{-0.1cm}v_{\a}
\hspace{-0.1cm}-\hspace{-0.1cm}\eta)}\no\\
&&~~~~~~=\prod_{k\neq
\a}^M\frac{\sin(v_{\a}+v_k+2\eta)\sin(v_{\a}-v_k+\eta)}
{\sin(v_{\a}+v_k)\sin(v_{\a}-v_k-\eta)}\no\\
&&~~~~~~~~~~\times\prod_{k=1}^{2M}\frac{\sin(v_{\a}+z_k)\sin(v_{\a}-z_k)}
{\sin(v_{\a}+z_k+\eta)\sin(v_{\a}-z_k+\eta)},~~\a=1,\cdots,M,
\label{BA-D-1}\eea the Bethe state $|v_1,\cdots,v_M\rangle^{(1)}$
becomes the eigenstate of the transfer matrix with eigenvalue
$\L^{(1)}(u)$  given by \cite{Yan04-1} \bea
&&\L^{(1)}(u)=\frac{\sin(\l_2+\bar\xi-u)\sin(\l_1+\bar\xi+u)\sin(\l_1+\xi-u)\sin(2u+2\eta)}
{\sin(\l_2+\bar\xi-u-\eta)\sin(\l_1+\bar\xi-u-\eta)\sin(\l_1+\xi+u)\sin(2u+\eta)}\no\\
&&~~~~~~~~~~~~~~~~~~\times\prod_{k=1}^M\frac{\sin(u+v_k)\sin(u-v_k-\eta)}
{\sin(u+v_k+\eta)\sin(u-v_k)}\no\\
&&~~~~~~+\frac{\sin(\l_2+\bar\xi+u+\eta)\sin(\l_1+\xi+u+\eta)\sin(\l_2+\xi-u-\eta)\sin
2u}
{\sin(\l_2+\bar\xi-u-\eta)\sin(\l_1+\xi+u)\sin(\l_2+\xi+u)\sin(2u+\eta)}\no\\
&&~~~~~~~~~~~~~~~~~~\times\prod_{k=1}^M\frac{\sin(u+v_k+2\eta)\sin(u-v_k+\eta)}
{\sin(u+v_k+\eta)\sin(u-v_k)}\no\\
&&~~~~~~~~~~~~~~~~~~\times\prod_{k=1}^{2M}\frac{\sin(u+z_k)\sin(u-z_k)}
{\sin(u+z_k+\eta)\sin(u-z_k+\eta)}.\label{Eigenfuction-D-1}
 \eea

\section{Second reference state and associated Bethe states}
\setcounter{equation}{0} Let us introduce the second reference
state $|\O^{(2)}(\l)\rangle$, \bea \hspace{-1.2truecm}
|\O^{(2)}(\l)\rangle&=&\phi_{\l-(N-1)\eta\hat{2},\l-N\eta\hat{2}}(-z_1)\otimes
\phi_{\l-(N-2)\eta\hat{2},\l-(N-1)\eta\hat{2}}(-z_{2})\cdots\otimes
\phi_{\l,\l-\eta\hat{2}}(-z_N),\label{Vac-2}\eea and the
associated operators $\A^{(2)}$ and $\D^{(2)}$ which are linear
combinations of $\{\T(m|u)^i_i\}$,\bea
&&\A^{(2)}(m|u)=\frac{\sin(m_{21}+\eta)}{\sin(m_{21})}
\{\T(m|u)^1_1-R(2u,m+\eta\hat{2})^{1\,2}_{2\,1}\D^{(2)}(m|u)\},
\label{Def-AB-2}\\
&&\D^{(2)}(m|u)=\T(m|u)^2_2. \label{Def-D-2}\eea Using the
technique developed in \cite{Yan04}, after tedious calculations,
we find that the state $|\O^{(2)}(\l)\rangle$ given by
(\ref{Vac-2}) is {\it exactly} the reference state in the
following sense, \bea
&&\A^{(2)}(\l-N\eta\hat{2}|u)|\O^{(2)}(\l)\rangle=\frac{\sin
2u\sin(\l_2+\xi+u+\eta)\sin(\l_1+\xi-u-\eta)}
{\sin(2u+\eta)\sin(\l_2+\xi+u)\sin(\l_1+\xi+u)}\no\\
&&~~~~~~~~~~~~~~~~~~~~~~~~~~~~\times\lt\{\prod_{k=1}^N
\frac{\sin(u+z_k)\sin(u-z_k)}{\sin(u+z_k+\eta)\sin(u-z_k+\eta)}\rt\}
|\O^{(2)}(\l)\rangle,\label{A-2}\\
&&\D^{(2)}(\l-N\eta\hat{2}|u)|\O^{(2)}(\l)\rangle=
\frac{\sin(\l_2+\xi-u)}{\sin(\l_2+\xi+u)}
|\O^{(2)}(\l)\rangle,\label{D-2}\\
&&\B(\l-N\eta\hat{2}|u)|\O^{(2)}(\l)\rangle=0,\label{Ann-1}\\
&&\C(\l-N\eta\hat{2}|u)|\O^{(2)}(\l)\rangle\neq 0. \eea Then  the
second set of Bethe states can be constructed by applying the
``creation operator" $\C$ on the reference state
$|\O^{(2)}(\l)\rangle$ (c.f. (\ref{Bethe-state})) \bea
|v_1,\cdots,v_M\rangle^{(2)}=\C(\l-2\eta\hat{2}|v_1)
\C(\l-4\eta\hat{2}|v_2)\cdots\C(\l-2M\eta\hat{2}|v_M)|\O^{(2)}(\l)\rangle.
\label{Bethe-state-2}\eea One may check that the transfer matrix
$\t(u)$ (\ref{trans}) is a linear combination of the operators
$\A^{(2)}$ and $\D^{(2)}$  \bea
&&\t(u)=\frac{\sin(\l_1+\bar{\xi}+u+\eta)}
{\sin(\l_1+\bar{\xi}-u-\eta)}\A^{(2)}(\l|u)
\no\\
&&~~~~~~~~~~~~~~~~~~+
\frac{\sin(\l_1+\bar{\xi}-u)\sin(\l_2+\bar{\xi}+u)\sin(2u+2\eta)}
{\sin(\l_1+\bar{\xi}-u-\eta)\sin(\l_2+\bar{\xi}-u-\eta)\sin(2u+\eta)}
\D^{(2)}(\l|u). \label{Exp-trans-2}\eea Carrying out the
generalized Bethe Ansatz \cite{Yan04-1,Yan04}, we finally find
that if the parameters $\{v_k\}$ satisfy the second Bethe Ansatz
equations (comparing with the first ones (\ref{BA-D-1})), \bea
&&\hspace{-0.1cm}\frac
{\sin(\l_1+\xi+v_{\a})\sin(\l_1+\bar\xi-v_{\a})
\sin(\l_2+\bar\xi+v_{\a})\sin(\l_2+\xi-v_{\a})}
{\sin(\l_1\hspace{-0.1cm}+\hspace{-0.1cm}
\bar\xi\hspace{-0.1cm}+\hspace{-0.1cm}v_{\a}
\hspace{-0.1cm}+\hspace{-0.1cm}\eta)
\sin(\l_1\hspace{-0.1cm}+\hspace{-0.1cm}\xi\hspace{-0.1cm}-\hspace{-0.1cm}v_{\a}
\hspace{-0.1cm}-\hspace{-0.1cm}\eta)
\sin(\l_2\hspace{-0.1cm}+\hspace{-0.1cm}\xi\hspace{-0.1cm}+\hspace{-0.1cm}
v_{\a}\hspace{-0.1cm}+\hspace{-0.1cm}\eta)
\sin(\l_2\hspace{-0.1cm}+\hspace{-0.1cm}\bar\xi\hspace{-0.1cm}-\hspace{-0.1cm}v_{\a}
\hspace{-0.1cm}-\hspace{-0.1cm}\eta)}\no\\
&&~~~~~~=\prod_{k\neq
\a}^M\frac{\sin(v_{\a}+v_k+2\eta)\sin(v_{\a}-v_k+\eta)}
{\sin(v_{\a}+v_k)\sin(v_{\a}-v_k-\eta)}\no\\
&&~~~~~~~~~~\times\prod_{k=1}^{2M}\frac{\sin(v_{\a}+z_k)\sin(v_{\a}-z_k)}
{\sin(v_{\a}+z_k+\eta)\sin(v_{\a}-z_k+\eta)},~~\a=1,\cdots,M,
\label{BA-D-2}\eea the Bethe states $|v_1,\cdots,v_M\rangle^{(2)}$
yield the second set of the eigenstates of the transfer matrix
with the eigenvalues, \bea
&&\L^{(2)}(u)=\frac{\sin(2u+2\eta)\sin(\l_1+\bar\xi-u)\sin(\l_2+\bar\xi+u)\sin(\l_2+\xi-u)}
{\sin(2u+\eta)\sin(\l_1+\bar\xi-u-\eta)\sin(\l_2+\bar\xi-u-\eta)\sin(\l_2+\xi+u)}\no\\
&&~~~~~~~~~~~~~~~~~~\times\prod_{k=1}^M\frac{\sin(u+v_k)\sin(u-v_k-\eta)}
{\sin(u+v_k+\eta)\sin(u-v_k)}\no\\
&&~~~~~~+\frac{\sin(2u)\sin(\l_1+\bar\xi+u+\eta)
\sin(\l_2+\xi+u+\eta)\sin(\l_1+\xi-u-\eta)}
{\sin(2u+\eta)\sin(\l_1+\bar\xi-u-\eta)\sin(\l_2+\xi+u)\sin(\l_1+\xi+u)}\no\\
&&~~~~~~~~~~~~~~~~~~\times\prod_{k=1}^M\frac{\sin(u+v_k+2\eta)\sin(u-v_k+\eta)}
{\sin(u+v_k+\eta)\sin(u-v_k)}\no\\
&&~~~~~~~~~~~~~~~~~~\times\prod_{k=1}^{2M}\frac{\sin(u+z_k)\sin(u-z_k)}
{\sin(u+z_k+\eta)\sin(u-z_k+\eta)}.\label{Eigenfuction-D-2}
 \eea Note that the normalizations adopted in this paper for the R- and
K-matrices are different from those used in  \cite{Yan06}. After
rescaling an overall factor
$\sin(\l_1+\bar{\xi}-u-\eta)\sin(\l_2+\bar{\xi}-u-\eta)\sin(\l_1+\xi+u)
\sin(\l_2+\xi+u)\prod_{k=1}^N\sin(u+z_k+\eta)\sin(u-z_k+\eta)$ and
setting all inhomogeneous parameters $z_k=0$, our two eigenvalues
$\{\L^{(i)}(u)\}$ recover those in \cite{Yan06}. Therefore two
sets Bethe states $\{|v_1,\cdots,v_M\rangle^{(i)}\}$
(\ref{Bethe-state}) and (\ref{Bethe-state-2}) together constitute
the complete eigenstates of the transfer matrix $\t(u)$
(\ref{trans}).

\section{Results for the Gaudin model}
\setcounter{equation}{0} In order to study the associated Gaudin
model, we need further restrict $\bar{\xi}=\xi$ \cite{Yan04-1}.
Following \cite{Hik95,Skl96} one can introduce the corresponding
Gaudin operators $\{H_j\}$ by expanding the double-row transfer
matrix $\t(u)$ (\ref{trans}) at the point $u=z_j$ around $\eta=0$:
\bea &&\t(z_j)={\rm id}+\eta H_j+O(\eta^2),\quad {\rm with }\,
H_j=\frac{\partial}{\partial
\eta}\t(z_j)|_{\eta=0},~~j=1,\cdots,N, \label{trans-2}\eea where
\bea &&H_j=\G_j(z_j)+\sum_{k\neq
j}^{2M}\frac{1}{\sin(z_j-z_k)}\lt\{\s^+_k\s^-_j+\s^-_k\s^+_j
+\cos(z_j-z_k)\frac{\s^z_k\s^z_j-1}{2}\rt\}\no\\
&&~~+\sum_{k\neq
j}^{2M}\frac{K_j^{-1}(z_j)}{\sin(z_j+z_k)}\lt\{\s^+_j\s^-_k+\s^-_j\s^+_k
+\cos(z_j+z_k)\frac{\s^z_j\s^z_k-1}{2}\rt\}K_j(z_j),\label{Ham}
\eea  where $\G_j(u)=\frac{\partial}{\partial
\eta}\{\bar{K}_j(u)\}|_{\eta=0}K_j(u)$, $j=1,\cdots,N,$ with
$\bar{K}_j(u)=tr_0\lt\{K^+_0(u)\R_{0j}(2u)R_{0j}(0)\rt\}$. The
commutativity of the transfer matrices $\{\t(z_j)\}$ for a generic
$\eta$ implies $ [H_j,H_k]=0$, for $i,j=1,\cdots,N$. Thus the
Gaudin system defined by (\ref{Ham}) is integrable. Moreover the
relation (\ref{trans-2}) between $\{H_j\}$ and $\{\t(z_j)\}$
enable us to extract the eigenstates of the Gaudin operators and
the corresponding eigenvalues from the results obtained in last
section.

Let us introduce states $|\bar{\O}^{(i)}(\l)\rangle$,\bea
|\bar{\O}^{(i)}(\l)\rangle=
\lt(\begin{array}{c}e^{i(z_1-2\l_{i})}\\1\end{array}\rt)
\otimes\cdots\otimes
\lt(\begin{array}{c}e^{i(z_N-2\l_{i})}\\1\end{array}\rt), \quad
i=1,2.\eea These states can be obtained from the reference states
$|\O^{(i)}(\l)\rangle$ by taking the limit:
$|\bar{\O}^{(i)}(\l)\rangle=\lim_{\eta\rightarrow
0}|\O^{(i)}(\l)\rangle$. Let us introduce a matrix $C(u)\in {\rm
End}(V)$ associated with the intertwiner vector $\phi$
(\ref{Intvect}) \bea
C(u)=\lt(\begin{array}{cc}e^{-i(u+2\l_1)}&e^{-i(u+2\l_2)}
\\1&1\end{array}\rt),\label{Matrix-in}\eea
and the corresponding gauged Pauli operator $\s^{\pm}(u)\in {\rm
End}(V)$: $\s^{\pm}(u)=C(u)\s^{\pm}C(u)^{-1}$. Then we can
construct  states $\Psi^{(i)}(x_1,\cdots,x_M)$: \bea
&&\Psi^{(1)}(x_1,\cdots,x_M)=\prod_{\a=1}^M\lt(\sum_{k=1}^{2M}\lt\{
\frac{\sin(\l_1+\xi-x_{\a})\sin(x_{\a}-z_k+\l_{12})}
{\sin(\l_1+\xi+x_{\a})\sin(x_{\a}-z_k)}\rt.\rt.\no\\
&&~~~~~~~~~~~~~~~~~~~~-
\lt.\lt.\frac{\sin(\l_2+\xi-x_{\a})\sin(x_{\a}+z_k-\l_{12})}
{\sin(\l_2+\xi+x_{\a})\sin(x_{\a}+z_k)} \rt\}\s_k^-(-z_k)\rt)
|\bar{\O}^{(1)}(\l)\rangle,\label{BAS-1}\\
&&\Psi^{(2)}(x_1,\cdots,x_M)=\prod_{\a=1}^M\lt(\sum_{k=1}^{2M}\lt\{
\frac{\sin(\l_2+\xi-x_{\a})\sin(x_{\a}-z_k-\l_{12})}
{\sin(\l_2+\xi+x_{\a})\sin(x_{\a}-z_k)}\rt.\rt.\no\\
&&~~~~~~~~~~~~~~~~~~~~-
\lt.\lt.\frac{\sin(\l_1+\xi-x_{\a})\sin(x_{\a}+z_k+\l_{12})}
{\sin(\l_1+\xi+x_{\a})\sin(x_{\a}+z_k)} \rt\}\s_k^+(-z_k)\rt)
|\bar{\O}^{(2)}(\l)\rangle.\label{BAS-2}\eea

Noting the relations (\ref{trans-2}) and using the same method as
in \cite{Yan04-1}, we find that if the parameters $\{x_k\}$
satisfy the following Bethe Ansatz equations \bea
&&\sum_{j=1}^2\frac{1}{\sin(\l_j+\xi-x_{\a})
\sin(\l_j+\xi+x_{\a})}+\sum_{k=1}^{2M} \frac{1}{\sin(x_{\a}+z_k)
\sin(x_{\a}-z_k)}\no\\
&&~~~~~~~~~~~~=2\sum_{k\neq \a}^M \frac{1}{\sin(x_{\a}+x_k)
\sin(x_{\a}-x_k)},~~\a=1,\cdots,M,\label{BAE-1}\eea the two sets
of  states $\Psi^{(i)}(x_1,\cdots,x_M)$ constitute the {\it
entire} eigenstates of the Gaudin operators \bea
&&H_j\Psi^{(i)}(x_1,\cdots,x_M)=E_j\Psi^{(i)}(x_1,\cdots,x_M),
\quad i=1,2.\label{Relation-BA}\eea The functions $E_j$ is \bea
&&E_j=\cot 2z_j+\sum_{j=1}^2\cot(\l_j+\xi-z_j)+\sum_{k=1}^M
\frac{\sin 2z_j}{\sin(x_k-z_j)\sin(x_k+z_j)}.\label{Eig-1} \eea
\section{Conclusions}
\label{Con} \setcounter{equation}{0} We have studied the second
reference state of the open XXZ spin chain with non-diagonal
boundary term, which leads to the second set of Bethe state
(\ref{Bethe-state-2}). These Bethe states give rise to the
corresponding Bethe Ansatz equations (\ref{BA-D-2}) and
eigenvalues (\ref{Eigenfuction-D-2}) proposed in
\cite{Nep03,Yan06} by the functional Bethe Ansatz method. In the
quasi-classical limit, two sets of Bethe states (\ref{BAS-1}) and
(\ref{BAS-2}) constitute the complete eigenstates of the
associated Gaudin model.

Very recently, an exact solution of the eigenvalue of the transfer
matrix of open XXZ spin chain for arbitrary boundary parameters
was proposed by functional Bethe Ansatz \cite{Mur06} and by
representation of q-Onsager algebra \cite{Bas07}. It would be
interesting to rederive their results in the framework of
algebraic Bethe Ansatz. Moreover, such structure of multiply
reference states found here also appears in open spin chains
associated with higher rank algebras \cite{Yan07}.

\section*{Acknowledgements}
The financial support from  Australian Research Council is
gratefully acknowledged.



\begin{thebibliography}{99}
\bibitem{Gau71} M. Gaudin, {\it Phys. Rev. } {\bf A 4} (1971), 386.
\bibitem{Alc87} F.\,C. Alcaraz, M.\,N. Barber, M.\,T. Batchelor,
R. \,J. Baxter and G.\,R.\,W. Quispel, {\it J. Phys. \/} {\bf A
20} (1987), 6397.
\bibitem{Skl88} E.\,K. Sklyanin, {\it J. Phys. \/} {\bf A 21}
(1988), 2375.
\bibitem{Nep04} R.\,I. Nepomechie,
{\it J. Stat. Phys.\/} {\bf 111} (2003), 1363; {\it J. Phys.\/}
{\bf A 37} (2004), 433.
\bibitem{Cao03} J. Cao, H.\,-Q. Lin, K.\,-J. Shi and Y. Wang, {\it
Nucl. Phys.\/} {\bf B 663} (2003), 487.
\bibitem{Yan04-1} W.\,-L. Yang, Y.\,-Z. Zhang and M. Gould, {\it
Nucl. Phys.\/} {\bf B 698} (2004), 312.
\bibitem{Gie05} J. de Gier, A. Nichols, P. Pyatov and V.
Rittenberg, {\it Nucl. Phys.\/} {\bf B 729} (2005), 387.
\bibitem{Doi06} A. Doikou and P.\,P. Martin, {\it J. Stat.
Mech.\/} {\bf P06004} (2006).
\bibitem{Yan06} W.\,-L. Yang, R.\,I. Nepomechie and Y.\,-Z. Zhang,
{\it Phys. Lett.\/} {\bf B 633} (2006), 664.
\bibitem{Nep03} R.\,I. Nepomechie and F. Ravanini, {\it J.
Phys.\/} {\bf A 36} (2003), 11391; Addendum, {\it J. Phys. \/}
{\bf A 37} (2004), 1945.
\bibitem{Kor93} V.\,E. Korepin, N.\,M. Bogoliubov and A.\,G. Izergin,
{\it Quantum Inverse Scattering Method and correlation
Function\/}, Cambridge Univ. Press, Cambridge, 1993.
\bibitem{Veg93} H.\,J. de Vega and A. Gonzalez-Ruiz, {\it J. Phys.
\/} {\bf A 26} (1993), L519.
\bibitem{Gho94} S. Ghoshal and A.\,B. Zamolodchikov, {\it Int. J.
Mod. Phys.\/} {\bf A 9} (1994), 3841.
\bibitem{Bax82} R.\,J. Baxter, {\it Exactly solved models in
statistical mechanics}, Academic Press, New York, 1982.
\bibitem{Yan04} W.\,-L. Yang and R. Sasaki,
{\it Nucl. Phys.\/} {\bf B 679} (2004), 495£» {\it J. Math.
Phys.\/} {\bf 45\/} (2004), 4301.
\bibitem{Hik95} K. Hikami, {\it J. Phys.\/} {\bf A 28} (1995),
4997.
\bibitem{Skl96} E.\,K. Sklyanin and T. Takebe, {\it Phys. Lett.\/}
{\bf A 219} (1996), 217.
\bibitem{Mur06} R. Murgan, R.\,I. Nepomechie and C. Shi, {\it J.
Stat. Mech.\/} {\bf P08006}, (2006).
\bibitem{Bas07} P. Baseilhac and K. Koizumi, {\tt hep-th/0703106}.
\bibitem{Yan07} W.\,-L. Yang and Y.\,-Z. Zhang, in preparation.

\end{thebibliography}
\end{document}